\documentclass{svproc}
\usepackage{url}
\usepackage{epsfig}  
\usepackage{graphicx}\usepackage{hyperref}
\usepackage{color}
\usepackage{float}
\usepackage{amsfonts}
\usepackage{amsmath}
\usepackage{slashed}
\def\UrlFont{\rmfamily}
\begin{document}
\mainmatter              
%

\title{$B^*_s\rightarrow l^+l^-$ decays in light of recent $B$ anomalies}
\titlerunning{$B^*_s\rightarrow l^+l^-$ decays in light of recent $B$ anomalies}  
%
\author{Suman Kumbhakar\inst{1} \and Jyoti Saini\inst{2}}
\authorrunning{Suman Kumbhakar and Jyoti Saini} 
%
\tocauthor{Jyoti Saini}
\institute{Indian Institute of Technology, Bombay\\ \email{suman@phy.iitb.ac.in}
\and
Indian Institute of Technology, Jodhpur}

\maketitle              

\begin{abstract}
Some of the recent measurements in the neutral current  sector $b\rightarrow s l^+l^-$ ($l=e$ or $\mu$) as well as in the charged current sector $b \rightarrow c \tau \bar{\nu}$ show significant deviations from their Standard Model predictions. It has been shown that two different new physics solutions, in the form of vector and/or axial vector, can explain all the anomalies in $b\rightarrow s l^+l^-$ sector. 
We show that the muon longitudinal polarization asymmetry in $B^*_s\rightarrow \mu^+\,\mu^-$ decay is a good discriminant between the two solutions if it can be measured to a precision of $\sim 10\%$, provided the new physics Wilson coefficients are real. We also investigate the potential impact of $b \rightarrow c \tau \bar{\nu}$ anomalies on $B_s^* \rightarrow \tau^+ \tau^-$ decay. We consider a model where the new physics contributions to these two transitions are strongly correlated. We find that two orders of magnitude enhancement in the branching ratio of $B^*_s\rightarrow \tau^+\,\tau^-$ is allowed by the present $b \rightarrow c \tau \bar{\nu}$ data.
\end{abstract}
%
\section{Introduction} 
The recent anomalies in the charged current (CC) transition $b\rightarrow c \tau\bar{\nu}$ and in the flavor changing neutral current (FCNC) transitions $b\rightarrow sl^+l^-$ ($l=e$ or $\mu$) provide tantalizing hints of physics beyond Standard Model (SM). In the SM, the above CC transition occurs at tree level whereas the FCNC transitions occur only at loop level. 

Some of the anomalies in $b\rightarrow sl^+l^-$ sector are: angular observables in $B\rightarrow K^* \mu^+ \mu^-$ \cite{Aaij2013,Aaij2016,Abdesselam} particularly $P^{'}_5$ in $4.3$-$8.68$ GeV$^2$ bin, the branching ratio of $B_s\rightarrow \phi \mu^+ \mu^-$ and the corresponding angular observables~\cite{Aaij2013JHEP,Aaij20171509}, the flavor ratio $R_K \equiv \Gamma(B^+ \rightarrow K^+ \mu^+ \mu^-)/\Gamma(B^+ \rightarrow K^+ e^+ e^-)$ in $1.0 \leq q^2 \leq 6.0$ GeV$^2$~\cite{Aaij2014}, the ratio $R_{K^*} \equiv \Gamma(B^0 \rightarrow K^{*0} \mu^+ \mu^-)/\Gamma(B^0 \rightarrow K^{*0} e^+ e^-)$ in two different $q^2$ ranges, $(0.045 \leq q^2 \leq 1.1$ GeV$^2)$ (low $q^2$) and $(1.1 \leq q^2 \leq 6.0$ GeV$^2)$ (central $q^2$)~\cite{Aaij2017}. In Moriond'19, the Belle collaboration has published their first measurements of $R_{K^*}$ in both $B^0$ and $B^+$ decays. These measurements are reported in multiple $q^2$ bins and have comparatively large uncertainties~\cite{Abdesselam:2019wac}. Further, LHCb collaboration updated the value of $R_K$ in Moriond'19 \cite{Aaij:2019wad}. 
After Moriond'19, refs.~\cite{Alok:2019ufo,Alguero:2019ptt} performed a global fit to identify the Lorentz structure of new physics (NP) which can account for all  anomalies in $b\rightarrow s \mu^+ \mu^-$ sector.  In 1D scenario, there are two distinct solutions, one with the operator of the form $(\bar{s}\gamma^{\alpha}P_Lb)(\bar{\mu}\gamma_{\alpha}\mu)$ and the other whose operator is a linear combination of $(\bar{s}\gamma^{\alpha}P_Lb)(\bar{\mu}\gamma_{\alpha}\mu)$ and $(\bar{s}\gamma^{\alpha}P_Lb)(\bar{\mu}\gamma_{\alpha}\gamma_5\mu)$.

It is interesting to look for new observables in the $b \rightarrow s \mu^+ \mu^-$ sector in order to (a) find additional evidence for the existence of NP and (b) to discriminate between the two NP solutions. 
The branching ratio of $B_s^*\rightarrow \mu^+\mu^- $ is one such observable which is yet to be measured. In the SM, this decay mode is not subject to helicity suppression~\cite{Grinstein1509}, unlike $B_s\rightarrow \mu^+\mu^-$. A model independent analysis of this decay was performed in ref.~\cite{Kumar:2017xgl} to identify the NP operators which can lead to a large enhancement of its branching ratio. It was found that such an enhancement is not possible due to the constraints from the present $b\rightarrow s\mu^+\mu^-$ data.  
In this work, we consider the longitudinal polarization asymmetry of muon in $B^*_s \rightarrow \mu^+ \mu^-$ decay, $\mathcal{A}_{LP}(\mu)$. This asymmetry is theoretically clean because it has a very mild dependence on the decay constants unlike the branching ratio. We first calculate the SM prediction of  $\mathcal{A}_{LP}(\mu)$ and then study its sensitivity to the NP solutions. 

 On the other hand, the discrepancies in the CC $b\rightarrow c \tau \bar{\nu}$ transition are: the ratios
$R_{D^{(*)}} = \Gamma(B\rightarrow D^{(*)}\,\tau\,\bar{\nu})/ \Gamma(B\rightarrow D^{(*)}\, \{e/\mu\} \, \bar{\nu})$~\cite{HFAG:2017avg},  $R_{J/ \psi} = \mathcal{B}(B\rightarrow J/ \psi\,\tau\,\bar{\nu})/\mathcal{B}(B \rightarrow J/ \psi \mu\, \bar{\nu})$~\cite{Aaij:2017tyk}. 
Refs.~\cite{Alok:2017qsi,Alok:2018uft,Alok:2019uqc} identified the allowed NP solutions which can explain all anomalies in the  $b \rightarrow c \tau \bar{\nu}$ sector and suggested methods to distinguish between various NP solutions. The NP WCs of these solutions are about $10\%$ of the SM values. Since this transition occurs at tree level in the SM, it is very likely that the NP operators also occur at tree level. 
In ref.~\cite{Capdevila:2017iqn}, a model is constructed where the tree level FCNC terms due to NP are significant for $b\rightarrow s\,\tau^+\,\tau^-$ but are suppressed for $b\rightarrow sl^+l^-$ where $l=e$ or $l=\mu$. The branching ratios for the decay modes such as $B\rightarrow K^{(*)}\tau^+\tau^-$, $B_s\rightarrow \tau^+\tau^-$ and $B_s\rightarrow \phi\tau^+\tau^-$ will have a large enhancement in this model~\cite{Capdevila:2017iqn}. In this work we study the effect of this NP on the branching ratio of $B_s^* \rightarrow \tau^+ \tau^-$ and the $\tau$ polarization asymmetry $\mathcal{A}_{LP}(\tau)$. 
\section{Longitudinal Polarization Asymmetry for $B_s^* \rightarrow l^+ l^-$ decay }
The decay  $B_s^* \rightarrow l^+\, l^-$ is induced by the quark level transition $ b\rightarrow s l^+ l^-$. In the SM the corresponding effective Hamiltonian is
\begin{align} \nonumber
\mathcal{H}_{SM} &= − \frac{4 G_F}{\sqrt{2} \pi} V_{ts}^* V_{tb} \bigg[ \sum_{i=1}^{6} C_i(\mu) O_i(\mu) + C_7 \frac{e}{16 \pi^2} [\overline{s} \sigma_{\mu \nu}(m_s P_L  + 
m_b P_R)b]F^{\mu \nu} \\ \nonumber & + C_9 \frac{\alpha_{em}}{4 \pi}(\overline{s} \gamma^{\mu} P_L b)(\overline{l} \gamma_{\mu} l) + C_{10} \frac{\alpha_{em}}{4 \pi} 
 (\overline{s} \gamma^{\mu} P_L b)(\overline{l} \gamma_{\mu} \gamma_5 l) \bigg],
\end{align}
where $G_F$ is the Fermi constant, $V_{ts}$ and $V_{tb}$ are the Cabibbo-Kobayashi-Maskawa (CKM) matrix elements and $P_{L,R} = (1 \mp \gamma^{5})/2$ are the projection operators. The effect of the operators $O_i,\,i=1-6,8 $ can be embedded in the redefined effective Wilson coefficients as $C_7(\mu)\rightarrow C^{eff}_7(\mu,q^2)$ and $C_9(\mu)\rightarrow C^{eff}_9(\mu,q^2)$. 
The form factor parameterization of the $B_s^* \rightarrow l^+\, l^-$ decay amplitudes are given in ref.~\cite{Grinstein1509}. These parameterization depend on the decay constants of $B^*_s$ meson  $f_{B^*_s}$ and $f_{B^*_s}^T$. 

As the NP solutions to the $b \rightarrow s l^+ l^-$ anomalies are in the form of vector and axial-vector operators, we consider the addition of these NP operators to the SM effective Hamiltonian of $b \rightarrow s l^+ l^-$. Scalar and pseudo-scalar NP operators do not contribute to $B_s^*\rightarrow l^+l^-$ decay because $\langle0 \vert\bar{s} b \vert B_s^{*}(p_{B_s^*},\epsilon) \rangle = \langle0 \vert\bar{s} \gamma_5 b \vert B_s^{*}(p_{B_s^*},\epsilon) \rangle = 0$. The effective Hamiltonian now takes the form
\begin{align}
\mathcal{H}_{eff}(b \rightarrow s l^+ l^-) &= \mathcal{H}_{SM} + \mathcal{H}_{VA} ,
\end{align}
where $\mathcal{H}_{VA}$ is 
\begin{align} \nonumber
\mathcal{H}_{VA} &= \frac{\alpha_{em}\,G_F}{\sqrt{2} \pi} V_{ts}^* V_{tb} \bigg[C_9^{NP}(\overline{s} \gamma^{\mu} P_L b)(\overline{l} \gamma_{\mu} l) + C_{10}^{NP} (\overline{s} \gamma^{\mu} P_L b)(\overline{l} \gamma_{\mu} \gamma_{5} l) \bigg]. 
\end{align}
Here $C^{NP}_{9(10)}$ are the NP Wilson coefficients.

We define the longitudinal polarization asymmetry for the final state leptons in $B_s^* \rightarrow l^+ l^-$ decay. The unit longitudinal polarization four-vector in the rest frame of the lepton ($l^{+}$ or $l^-$) is defined as
\begin{align}
\overline{s}_{l^{\pm}}^{\alpha}= \left(0, \pm\frac{\overrightarrow{p_{l}}}{|\overrightarrow{p_{l}}|}\right).
\end{align} 
In the dilepton rest frame  (which is also the rest frame of $B_s^* $ meson), these unit polarization vectors become
\begin{align}
s_{l^{\pm}}^{\alpha}= \left(\frac{|\overrightarrow{p_{l}}|}{m_{\l}}, \pm \frac{E_{l}}{m_{l}} \frac{\overrightarrow{p_{l}}}{|\overrightarrow{p_{l}}|}\right),
\end{align}
where $E_{l}$, $\overrightarrow{p_{l}}$ and $m_{l}$ are the energy, momentum and mass of the lepton ($l^{+}$ or $l^-$) respectively. We can define two longitudinal polarization asymmetries, $\mathcal{A}_{LP}^{+}$ for $l^+$ and $\mathcal{A}_{LP}^{-}$ for $l^-$, in the decay $B_s^* \rightarrow l^+\, l^-$ as~\cite{Handoko:prd65}
\begin{equation}
\mathcal{A}_{LP}^{\pm}=\frac{[\Gamma( s_{l^-}, s_{l^+}) + \Gamma(\mp s_{l^-}, \pm s_{l^+})]-[\Gamma(\pm s_{l^-}, \mp s_{l^+}) + \Gamma(-s_{l^-},-s_{l^+})]}{[\Gamma( s_{l^-}, s_{l^+}) + \Gamma(\mp s_{l^-}, \pm s_{l^+})]+[\Gamma(\pm s_{l^-}, \mp s_{l^+}) + \Gamma(-s_{l^-},-s_{l^+})]}.
\label{alp}
\end{equation}
Within this NP framework, the branching ratio and $\mathcal{A}_{LP}$ are obtained to be~\cite{Kumbhakar:2018uty}
\begin{eqnarray}
\mathcal{B}(B^*_s\rightarrow l^+l^-) &=& \frac{\alpha^2_{em}G^2_Ff^2_{B^*_s}m^3_{B^*_s}\tau_{B^*_s}}{96\pi^3}\vert V_{ts}V^*_{tb}\vert^2\sqrt{1-4m_l^2/m^2_{B^*_s}}\left[\left(1+\frac{2m^2_l}{m^2_{B^*_s}}\right)\left\vert C_9^{eff} \right. \right. \nonumber\\
 & & \left.\left. +\frac{2 m_b f_{B_s^*}^T}{m_{B_s^*} f_{B_s^*}}C_7^{eff}+C_9^{NP} \right\vert^2  +\left(1-\frac{4m^2_l}{m^2_{B^*_s}}\right)\vert C_{10}+ C_{10}^{NP}\vert^2\right],
\end{eqnarray}
\begin{align}
\mathcal{A}_{LP}^{\pm}\vert_{NP}= \mp \frac{ 2\sqrt{1-4 m_{l}^2/m^2_{B_s^*}}~Re\left[\left(C_9^{eff}+\frac{2 m_b f_{B_s^*}^T}{m_{B_s^*} f_{B_s^*}}C_7^{eff}+C_9^{NP} \right) \left(C_{10}+C_{10}^{NP}\right)^*\right]}{\left(1 + 2 m_{l}^2/m_{B_s^*}^2 \right) \left|C_9^{eff}+\frac{2 m_b f_{B_s^*}^T}{m_{B_s^*} f_{B_s^*}}C_7^{eff} +C_9^{NP}\right|^2 + \left(1 - 4 m_{l}^2/m_{B_s^*}^2\right) \left|C_{10 }+C_{10}^{NP}\right|^2}.
\label{eqNP}
\end{align}

\section{Results and Discussion}
\subsection{ $\mathcal{A}_{LP}(\mu)$ with NP solutions}

In this section we first calculate $\mathcal{A}_{LP}(\mu)$ for the $B_s^* \rightarrow \mu^+ \mu^-$ decay. The numerical inputs used for this calculation are $m_b=4.18\pm 0.03$ GeV, $m_{B^*_s}= 5415.4^{+1.8}_{-1.5}$ MeV~\cite{Patrignani:2016xqp}, $f^T_{B^*_s}/f_{B_s} = 0.95$~\cite{Grinstein1509} and $f_{B^*_s}/f_{B_s}= 0.953\pm 0.023$~\cite{Colquhoun:2015oha}. The SM prediction  is given in table~\ref{tab2}. 
The uncertainty in this prediction (about $0.03\%$) is much smaller than the uncertainty in the decay constants (about $2\%$), making it theoretically clean.

\begin{table*}[h]
	\centering 
	\tabcolsep 3pt
	\begin{tabular}{|c|c|c|c|}
		\hline\hline
		NP type  &  NP WCs & $\mathcal{B}(B^*_s\rightarrow \mu^+\mu^-)$ & $\mathcal{A}^+_{LP}(\mu)=-\mathcal{A}^-_{LP}(\mu)$  \\
		\hline
		SM & 0 & $(1.10\pm 0.60)\times 10^{-11}$ & $0.9955 \pm 0.0003$ \\ \hline
		(I) $C_9^{NP}(\mu\mu)$~~~~~~~~~~~~~ & $-1.07\pm 0.18$ & $(0.82\pm 0.50)\times 10^{-11}$ & $0.9145 \pm 0.0246$\\
		
		(II) $C_9^{NP}(\mu\mu)=-C_{10}^{NP}(\mu\mu)$  & $-0.52 \pm 0.09$ & $(0.80\pm 0.49)\times 10^{-11}$  &$0.9940 \pm0.0038$ \\
		\hline\hline
	\end{tabular}
	\caption{ Predictions of branching ratio and $\mathcal{A}_{LP}(\mu)$ for $B_s^* \rightarrow \mu^+ \mu^-$ decay. The values of NP WCs are taken from \cite{Alok:2019ufo}.}
	\label{tab2}
\end{table*}
From this table it is obvious that the prediction of $\mathcal{A}_{LP}(\mu)$ for the first solution deviates from the SM at the level of $3\sigma$ whereas, for the second solution, it is the same as that of the SM. Hence any large deviation in this asymmetry can only be due to the first NP solution. We also provide the predictions for $\mathcal{B}(B^*_s\rightarrow \mu^+\mu^-)$ in table~\ref{tab2}. It is clear that neither of the two solutions can be distinguished from each other or from the SM via the branching ratio.
\begin{figure*}[htbp]
	\begin{tabular}{cc}
		\includegraphics[width=63mm]{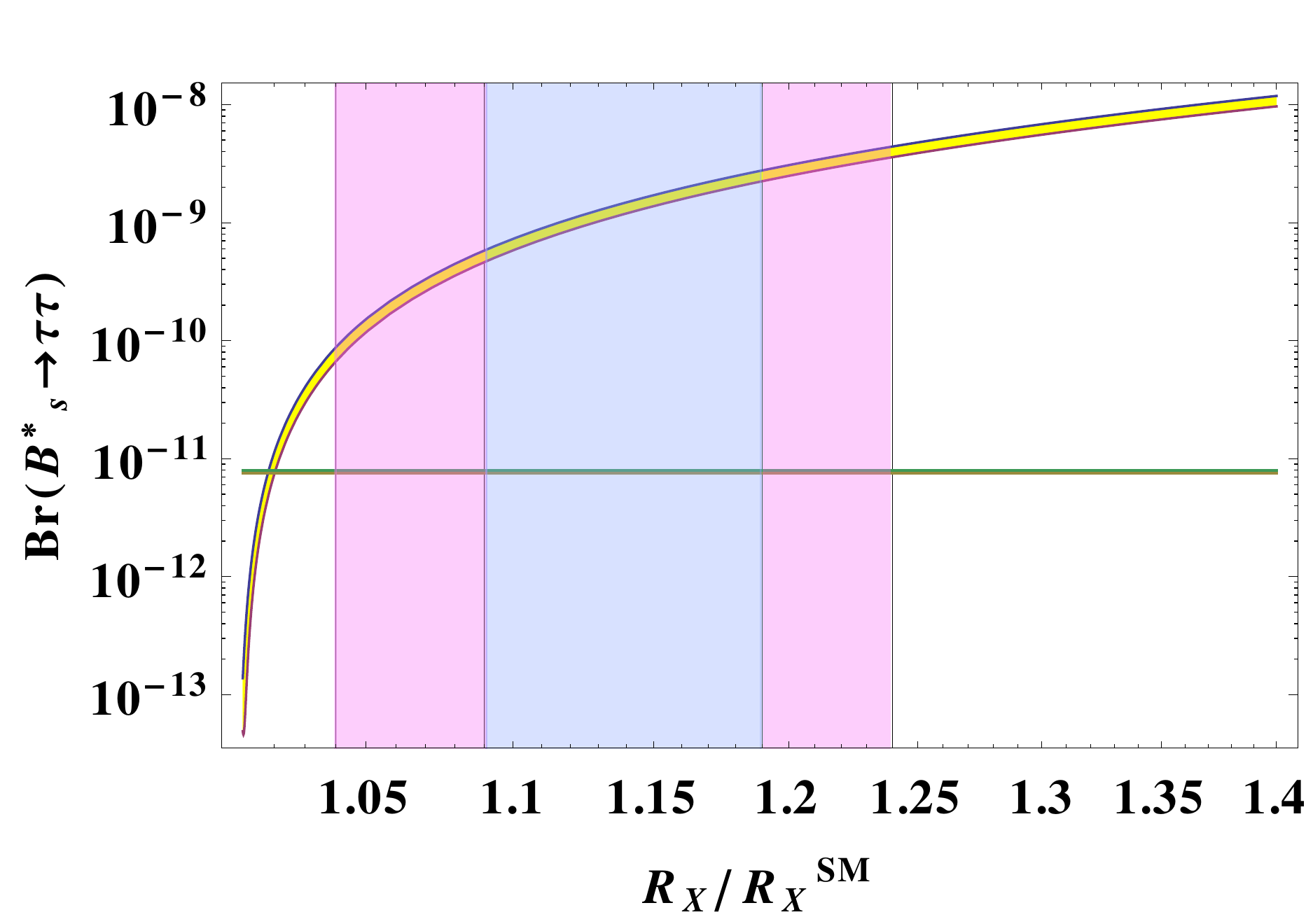}&
		\includegraphics[width=63mm]{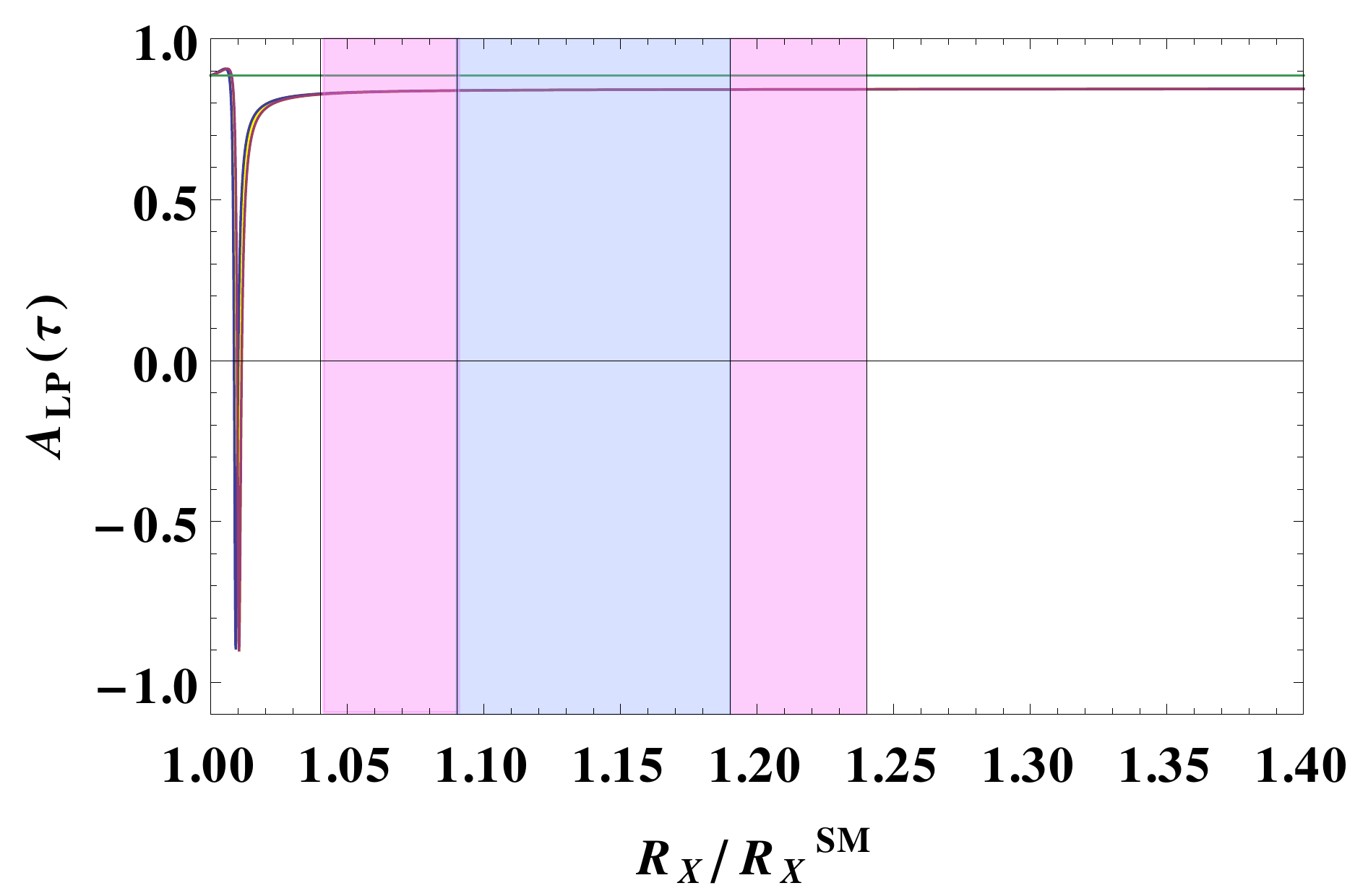}\\
	\end{tabular}
	\caption{Left and right panels correspond to $\mathcal{B}(B^*_s\rightarrow \tau^+\tau^-)$ and $\mathcal{A}_{LP}(\tau)$ respectively. In both panels the yellow band represents $1\sigma$ range of these observables. The $1\sigma$ and $2\sigma$ ranges of $R_X/R^{SM}_X$ are indicated by blue and pink bands respectively. The green horizontal line corresponds to the SM value.} 
	\label{fig1}
\end{figure*}

\subsection{Effect of NP in $B_s^* \rightarrow \tau^+ \tau^-$ }

 As mentioned in the introduction, anomalies are also observed in the $b\rightarrow c\tau\bar{\nu}$ transitions. An NP model, which can account for these anomalies, is likely to contain NP amplitude for $b\rightarrow s\tau^+\tau^-$ transition also. Hence the branching ratio of $B^*_s\rightarrow \tau^+\tau^-$ and $\tau$ longitudinal polarization asymmetry $\mathcal{A}_{LP}(\tau)$ will contain signatures of such NP. In the SM, the predictions for these quantities are:
$\mathcal{B}(B^*_s\rightarrow \tau^+\tau^-) = (6.87 \pm 4.23) \times 10^{-12}$ and
$\mathcal{A}^+_{LP}(\tau)|_{SM}=-\mathcal{A}^{-}_{LP}(\tau)|_{SM} = 0.8860\pm 0.0006$.

The authors of ref.~\cite{Capdevila:2017iqn} constructed a model of NP which accounts for the anomalies in $b\rightarrow c\tau \bar{\nu}$. This model contains tree level FCNC terms for $b\rightarrow s\,\tau^+\,\tau^-$ but not for $b\rightarrow s l^+ l^-$ ($l=e,\mu$).  The WCs for the $b\rightarrow s\tau^+\tau^-$ transition have the form
$C_{9}(\tau\tau) = C^{SM}_{9} - C^{NP}(\tau\tau)$ and 
$C_{10}(\tau\tau) = C^{SM}_{10} + C^{NP}(\tau\tau)$,
in this model, where 
\begin{equation}
C^{NP}(\tau\tau) = \frac{2\pi}{\alpha}\frac{V_{cb}}{V_{tb}V^*_{ts}}\left(\sqrt{\frac{R_X}{R^{SM}_X}}-1\right).
\end{equation}
The ratio $R_X/R^{SM}_X$ is the weighted average of current experimental values of $R_D$, $R_{D^*}$ and $R_{J/\psi}$. From the current world averages (after Moriond'19) of these quantities, we estimate this ratio to be $\simeq 1.14\pm 0.05$. This, in turn, leads to $C^{NP}(\tau\tau)\sim \mathcal{O}(100)$. Thus the NP contribution completely dominates the WCs and leads to greatly enhanced branching ratios for various $B$/$B_s$ meson decays involving $b\rightarrow s\,\tau^+\,\tau^-$ transition~\cite{Capdevila:2017iqn}. 
   
We calculate $\mathcal{B}(B^*_s\rightarrow \tau^+\tau^-)$ and $\mathcal{A}_{LP}(\tau)$ as a function of $R_X/R^{SM}_X$.
The plot of $\mathcal{B}(B^*_s\rightarrow \tau^+\tau^-)$ $vs.$ $R_X/R^{SM}_X$ is shown in left panel of fig.~\ref{fig1}. We note, from this plot, that  $\mathcal{B}(B_s^* \rightarrow \tau^+ \tau^-)$ can be enhanced up to $~10^{-9}$ which is about two orders of magnitude larger than the SM prediction. 
The plot of $\mathcal{A}_{LP}(\tau)$ $vs.$ $R_X/R^{SM}_X$ is shown in the right panel of fig.~\ref{fig1}. It can be seen that $\mathcal{A}_{LP}(\tau)$ is suppressed by about $5\%$ in comparison to its SM value.

After Moriond'19, the current world average of $R_{D^{(*)}}$ shows  less tension with the SM which leads to smaller values of $R_X/R_X^{SM}$. As long as this ratio is greater than $ 1.03$, the branching ratio of $B^*_s\rightarrow \tau^+\tau^-$ is enhanced by an order of magnitude at least. When $R_X/R^{SM}_X \sim 1.01$, $\mathcal{A}_{LP}(\tau)$ exhibits some very interesting behaviour. In this case, the tree level FCNC NP contribution is similar in magnitude to the SM contribution (which occurs only at loop level). Due to the interference between these two amplitudes, $\mathcal{A}_{LP}(\tau)$ changes sign and becomes almost ($-1$). Hence a measurement of this asymmetry provides an effective tool for the discovery of tree level FCNC amplitudes of this model~\cite{Capdevila:2017iqn} when their magnitude becomes quite small.

\section{Conclusions}
In this work we consider the ability of the muon longitudinal polarization asymmetry in $B_s^* \rightarrow \mu^+ \mu^-$ decay to distinguish
between the two NP solutions, $C_9^{NP}(\mu\mu)<0$ and $C_9^{NP}(\mu\mu)=-C_{10}^{NP}(\mu\mu)<0$, which can account for all the measurements in $b\rightarrow s l^+l^-$ sector. This observable is theoretically clean because it has only a very mild dependence on the decay constants. For the case of real NP WCs, we show that this asymmetry has the same value as the SM case for the second solution but is smaller by $\sim 10\%$ for the first solution. Hence, a measurement of this asymmetry to $\sim 10\%$ accuracy can distinguish between these two solutions.

Further, we study the impact of the anomalies in $b\rightarrow c\tau\bar{\nu}$ transitions on the branching ratio of $B_s^* \rightarrow \tau^+ \tau^-$ and $\mathcal{A}_{LP}(\tau)$. In ref.~\cite{Capdevila:2017iqn}, a model was constructed where tree level NP leads to both $b\rightarrow s\tau^+\tau^-$ and $b\rightarrow c\tau\bar{\nu}$ with moderately large NP couplings. Within this NP model, we find that the present data in $R_{D^{(*)},J/\psi}$ sector imply about two orders of magnitude enhancement in the branching ratio of $B_s^* \rightarrow \tau^+ \tau^-$ and a $5\%$ suppression  in $\mathcal{A}_{LP}(\tau)$ compared to their SM predictions. We also show that $\mathcal{A}_{LP}(\tau)$ undergoes drastic changes when the NP amplitude is similar in magnitude to the SM amplitude.



\begin{thebibliography}{10}

\bibitem{Aaij2013} 
R. Aaij et al. [LHCb Collaboration], Phys. Rev. Lett. 111, 191801 (2013).

\bibitem{Aaij2016} 
R. Aaij et al. [LHCb Collaboration], JHEP 1602, 104 (2016).

\bibitem{Abdesselam} 
A. Abdesselam et al. [Belle Collaboration], arXiv:1604.04042 [hep-ex].




\bibitem{Aaij2013JHEP} 
R. Aaij et al. [LHCb Collaboration], JHEP 1307, 084 (2013).

\bibitem{Aaij20171509} 
R. Aaij et al. [LHCb Collaboration], JHEP 1509, 179 (2015).



\bibitem{Aaij2014} 
R. Aaij  et al. [LHCb Collaboration], Phys. Rev. Lett. 113, 151601 (2014).
	
	


\bibitem{Aaij2017} 
R. Aaij et al. [LHCb Collaboration], JHEP 1708, 055 (2017).

\bibitem{Abdesselam:2019wac} 
  A.~Abdesselam {\it et al.} [Belle Collaboration],
  arXiv:1904.02440 [hep-ex].
  
  \bibitem{Aaij:2019wad}
  R.~Aaij {\it et al.} [LHCb Collaboration],
  Phys.\ Rev.\ Lett.\  {\bf 122} (2019) no.19,  191801

\bibitem{Alok:2019ufo}
  A.~K.~Alok, A.~Dighe, S.~Gangal and D.~Kumar,
  arXiv:1903.09617 [hep-ph].
  
\bibitem{Alguero:2019ptt} 
  M.~Algueró, B.~Capdevila, A.~Crivellin, S.~Descotes-Genon, P.~Masjuan, J.~Matias and J.~Virto,
  arXiv:1903.09578 [hep-ph].



\bibitem{Grinstein1509}
B.~Grinstein and J.~Martin Camalich,
Phys.\ Rev.\ Lett.\  {\bf 116} (2016) no.14,  141801.


\bibitem{Kumar:2017xgl}
D.~Kumar, J.~Saini, S.~Gangal and S.~B.~Das,
Phys.\ Rev.\ D {\bf 97} (2018) no.3,  035007.


  \bibitem{HFAG:2017avg}
  https://hflav-eos.web.cern.ch/hflav-eos/semi/spring19/html/RDsDsstar/RDRDs.html
  
  \bibitem{Aaij:2017tyk}
  R.~Aaij {\it et al.} [LHCb Collaboration],
  Phys.\ Rev.\ Lett.\  {\bf 120} (2018) no.12,  121801.
  


\bibitem{Alok:2017qsi}
  A.~K.~Alok, D.~Kumar, J.~Kumar, S.~Kumbhakar and S.~U.~Sankar,
  JHEP {\bf 1809} (2018) 152.
 
  
  \bibitem{Alok:2018uft}
  A.~K.~Alok, D.~Kumar, S.~Kumbhakar and S.~Uma Sankar,
  Phys.\ Lett.\ B {\bf 784} (2018) 16.
  
  \bibitem{Alok:2019uqc} 
  A.~K.~Alok, D.~Kumar, S.~Kumbhakar and S.~Uma Sankar,
  arXiv:1903.10486 [hep-ph].

 \bibitem{Capdevila:2017iqn}
  B.~Capdevila, A.~Crivellin, S.~Descotes-Genon, L.~Hofer and J.~Matias,
  Phys.\ Rev.\ Lett.\  {\bf 120} (2018) no.18,  181802.
  
\bibitem{Handoko:prd65}
L. T. Handoko, C. S. Kim and T. Yoshikawa, Phys. Rev. D 65, 077506 (2002).

  \bibitem{Kumbhakar:2018uty}
  S.~Kumbhakar and J.~Saini,
  Eur.\ Phys.\ J.\ C {\bf 79} (2019) no.5,  394.



\bibitem{Patrignani:2016xqp}
C.~Patrignani {\it et al.} [Particle Data Group],
Chin.\ Phys.\ C {\bf 40} (2016) no.10,  100001.



\bibitem{Colquhoun:2015oha}
B.~Colquhoun {\it et al.} [HPQCD Collaboration],
Phys.\ Rev.\ D {\bf 91} (2015) no.11,  114509.

  





\end{thebibliography}
\end{document}